\documentclass[aps,prd,preprintnumbers,superscriptaddress,endnote,nofootinbib,twocolumn]{revtex4-1}

\pdfoutput=1

\usepackage{graphicx,latexsym,amsfonts,amssymb,amsmath,slashed,array,dcolumn,feynmp}

\def\lsim{\mathrel{\raise.3ex\hbox{$<$\kern-.75em\lower1ex\hbox{$\sim$}}}}
\def\gsim{\mathrel{\raise.3ex\hbox{$>$\kern-.75em\lower1ex\hbox{$\sim$}}}}

\usepackage{xcolor}
\definecolor{red}{rgb}{1.0, 0, 0}

\newcommand{ \slashchar }[1]{\setbox0=\hbox{$#1$}   
   \dimen0=\wd0                                     
   \setbox1=\hbox{/} \dimen1=\wd1                   
   \ifdim\dimen0>\dimen1                            
      \rlap{\hbox to \dimen0{\hfil/\hfil}}          
      #1                                            
   \else                                            
      \rlap{\hbox to \dimen1{\hfil$#1$\hfil}}       
      /                                             
   \fi}                                             %

\newcommand{\gev}{\text{GeV}}

\newcommand{\ra}{\rightarrow}

\newcommand{\octet}{\Theta}
\newcommand{\scalar}{\phi}
\newcommand{\scalari}{\phi_i}
\newcommand{\Mi}{M_i}
\newcommand{\eps}{\epsilon}

\begin{document}

\title{Higgs Underproduction at the LHC}

\author{Bogdan A. Dobrescu}
\affiliation{Theoretical Physics Department, Fermilab, Batavia, IL 60510}

\author{Graham D. Kribs}
\affiliation{Theoretical Physics Department, Fermilab, Batavia, IL 60510}
\affiliation{Department of Physics, University of Oregon,  Eugene, OR 97403}

\author{Adam Martin}
\affiliation{Theoretical Physics Department, Fermilab, Batavia, IL 60510}

\preprint{FERMILAB-PUB-11-649-T}
\date{December 9, 2011; Revised January 2, 2012}

\begin{abstract}

We show that production of the Higgs boson through gluon-fusion 
may be suppressed in the presence of colored scalars.  
Substantial destructive interference between the top quark diagrams
and colored scalar diagrams is possible due to cancellations
between the real (and also imaginary) parts of the amplitudes.
As an example, we consider a color-octet scalar 
that has a negative, order one coupling to the Higgs doublet.
We find that gluon fusion can be suppressed by more than an 
order of magnitude when the scalar mass is below a few hundred GeV, 
while milder suppressions occur for larger scalar masses or 
smaller couplings.  Thus, the standard model extended with only 
one particle can evade the full range of present LHC exclusion limits 
on the Higgs mass.  The colored scalars, however, would be produced 
in pairs with a large rate at the LHC, leading to multi-jet final states 
to which the LHC experiments are now becoming sensitive.

\end{abstract}

\maketitle

\section{Introduction}
\label{sec:intro}

How effectively can new particles hide the Higgs boson from experiment?
With the ATLAS and CMS experiments at the Large Hadron Collider (LHC) 
having reached exclusion of the standard model Higgs boson
throughout a significant range of its mass \cite{atlascms}, 
this question has taken on 
heightened importance.  In this paper we demonstrate that 
gluon fusion \cite{HUTP-77/A084} -- the 
dominant production of Higgs boson at the LHC -- can be 
substantially reduced by one or more colored scalars with 
weak-scale mass and order-one couplings to the Higgs doublet.

Within the standard model, the overwhelmingly dominant contribution to 
Higgs production through gluon fusion comes from a top quark loop \cite{tasi}.
Beyond the standard model, there can be 1-loop contributions from 
particles that carry color and that also interact with the Higgs doublet. 
Fermions with renormalizable couplings to the Higgs doublet 
have contributions to the gluon fusion amplitude of the same sign 
as the top loop (e.g., a fourth generation \cite{fourggh}).
Large suppressions to gluon fusion thus appear to require 
some colored bosons.  

In this paper we consider the possible suppression of Higgs production
through gluon fusion in the presence of colored scalar fields.
One or more scalars $\scalar_i$ transforming under QCD can be 
coupled to the Higgs doublet through the renormalizable 
``Higgs portal'' interactions $-\kappa \scalar_i^\dagger\scalar_i H^\dagger H$ in the Lagrangian.  
We point out that the sign of the parameter $\kappa$ is not 
theoretically determined, so that for 
one choice, \emph{negative} $\kappa$, the scalar 
contribution interferes destructively with the top loop.
Examples of models with colored scalars where effects on 
Higgs production was discussed includes, for example, 
Refs.~\cite{susysuppression,Manohar:2006ga, Arnesen:2008fb, Ma:2011kc, arXiv:0709.4227,
Boughezal:2010ry}.

The reduction of gluon fusion has been noted previously in the
minimal supersymmetric standard model (MSSM), where squark loops 
may partially cancel the top loop for certain regions of 
parameter space \cite{susysuppression}.
In that case, the Higgs boson is already required to be rather light in the MSSM, 
in the mass region that is not yet ruled out by LHC and Tevatron data.  
Futhermore, supersymmetry requires
the assortment of colored superpartners 
that is being pushed to higher masses by the nonobservation results from 
the searches for supersymmetry at the LHC\@.  By contrast, 
we are interested in a more general scenario here, 
where the suppression of gluon fusion occurs for a wide range of 
Higgs masses, and the particle responsible for the suppression is 
harder to detect. The concrete example we study here is the standard model 
extended by an electroweak-singlet, color-octet real scalar~\cite{Bai:2010dj,HUTP-91-A009,Dobrescu:2007yp}.

Besides explicit, renormalizable models that include particles 
running in loops that suppress gluon fusion, one can imagine 
a strongly-coupled sector \cite{Manohar:2006gz,stronglyinteracting}
that generates the dimension-6 operator
$G_{\mu\nu} G^{\mu\nu} H^\dagger H/ (2\Lambda^2)$ in the Lagrangian
with the appropriate sign to cancel the top loop. 
In Ref.~\cite{Manohar:2006gz} it was shown that a coefficient of 
$-1$ for this operator 
leads to complete destructive interference with the standard model 
contribution for $\Lambda \simeq 3$~TeV\@. If this operator is 
generated by a 1-loop diagram involving a particle of mass $M$ 
and coupling of order one to the Higgs doublet, 
then a naive loop factor of $1/(4\pi)^2$ 
leads to a value $M\sim \Lambda/(4 \pi)$.
As we will see, the loop suppression is accidentally stronger, 
so that $M$ needs to be somewhat smaller than $\Lambda/(4 \pi)$. 
A detailed analysis is required to determine whether such 
light colored scalars are permitted by existing bounds from 
collider experiments.

We emphasize that this class of models leaves electroweak 
symmetry breaking unaffected. As a result,  the 
branching fractions of the Higgs boson remain
virtually unaffected throughout the Higgs mass range, especially when the
colored scalars are electroweak-singlet.
Only the decay width into gluon pairs is reduced 
when Higgs production through gluon fusion is suppressed,
but this decay is very hard to observe and its branching fraction 
is already smaller than about 9\%  for any Higgs mass allowed by LEP.
This is in contrast to models that modify the mechanism of electroweak symmetry 
breaking, which, not surprisingly, affect both Higgs production and decay, especially
in the light Higgs region \cite{arXiv:0801.4554}.

\bigskip
\section{Models of Underproduction}
\label{sec:noggh}
\setcounter{equation}{0}

The general class of models leading to modifications in 
Higgs production that we consider consist of the standard model extended
to include a set 
of real or complex  scalars $\scalari$ transforming under some
representations of the color $SU(3)$ group. 
The renormalizable interactions of the colored scalars are of
the form
\begin{eqnarray}
\mathcal{L}(\scalari) &=& 
        D_{\mu} \scalari^\dagger D^{\mu} \scalari 
      - \Mi^2 \, \scalari^\dagger \scalari \nonumber \\
& &{} - \kappa_{ij} \scalar_i^\dagger \scalar_j \, H^\dagger H 
      - \lambda_{ijkl} \scalar_i^\dagger \scalar_j \scalar_k^\dagger \scalar_l  
\label{eq:lagrangiangeneral}
\end{eqnarray}
where suitable color contractions are implicit (for $\lambda_{ijkl}$, this can result in several independent
interactions).  This set of interactions can be recast for real scalar
fields under the replacement $(\scalari, \scalari^\dagger) \ra 
(\scalari, \scalari)/\sqrt{2}$. 
Additional representation-dependent renormalizable interactions
are possible, such as 
$\eps_{\alpha\beta\gamma} \scalar_i^\alpha \scalar_j^\beta \scalar_k^\gamma$ for color triplets,
$d_{abc} \scalar_i^a \scalar_j^b \scalar_k^c$ for color octets, etc.

\begin{figure}[t]
\centerline{\includegraphics[width=0.27\textwidth]{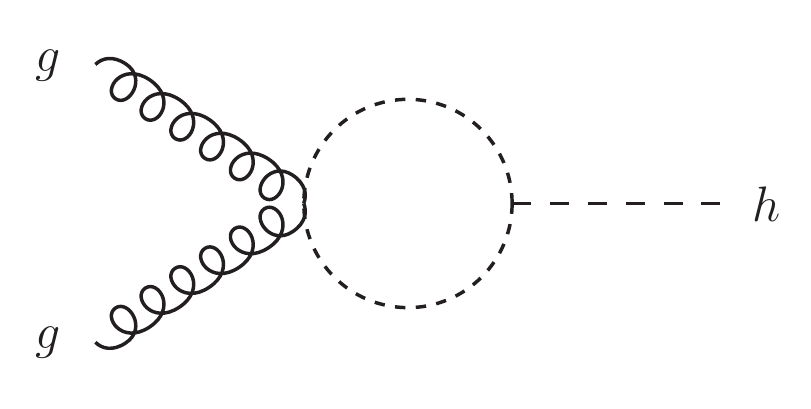}}
\centerline{\includegraphics[width=0.27\textwidth]{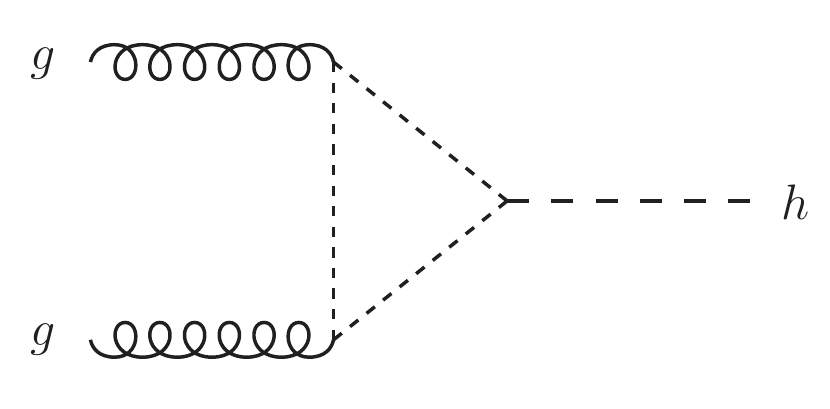}}
\caption{Feynman diagrams for scalar loop contributions 
to $gg \rightarrow h$.  A real scalar field has precisely these 
diagrams, while a complex field also has a third diagram that can be obtained from the second diagram by swapping 
the initial state gluons.}
\label{fig:feynggh}
\end{figure}

The Higgs portal interactions proportional to $\kappa$
are our primary interest.  Consider the effects of a single scalar field,
$\scalari$.  Expanding $H = (v + h)/\sqrt{2}$ gives the 
dimension-3 operator $\kappa_{ii} v \scalar_i^\dagger \scalar_i h$,
that leads to the 1-loop colored scalar contributions to
gluon fusion shown in Fig.~\ref{fig:feynggh}. 
For a complex scalar field, the diagrams in Fig.~\ref{fig:feynggh}
are added to the ``gluon-crossed'' triangle diagram to obtain a finite result, 
similar to the calculation of the top-quark loop.  For a real 
scalar field there is no gluon-crossed diagram, 
but the bubble diagram has a symmetry factor of $1/2$ such that the result is again finite.

These new physics contributions combine with the standard model 
contributions to the gluon fusion process.  
For production of a single on-shell Higgs in the narrow width approximation, 
the gluon fusion rate is proportional to the partial width of the 
Higgs boson into gluons,
\begin{eqnarray}
\hat{\sigma}(gg \ra h) &=& 
   \frac{\pi^2 \Gamma(h \ra gg)}{8 M_h} \delta(\hat{s} - M_h^2) \; .
\label{eq:productiongeneral}
\end{eqnarray}
At leading order, that partial width is
\begin{equation}
\Gamma(h \ra gg) = \frac{G_F \alpha_s^2 M_h^3}{64 \sqrt{2} \pi^3} 
\left|A_t(\tau_t) + 
      \sum_i c_i \kappa_{ii} \frac{v^2}{2 M_i^2} A_i(\tau_i)\right|^2
\label{eq:widthgeneral}
\end{equation}
where $c_i = C_A^i$ ($c_i = 2 C_A^i$) is equal to (twice) the quadratic 
Casimir of the QCD representation of the $i$th scalar in a real (complex)
representation.  Here $\tau_t \equiv M_h^2/(4 M_t^2)$ and 
$\tau_i \equiv M_h^2/(4 M_i^2)$ while $A_t$ and $A_i$ are
the contributions to the amplitude from top quark loops and scalar loops,
respectively. 
For the scalar contribution we obtain
\begin{eqnarray}
\tau_i A_i(\tau_i) &=& 
2 M_i^2 C_0(4 M_i^2 \tau_i; M_i) + 1 
\end{eqnarray}
in terms of the three-point Passarino-Veltman \cite{Passarino:1978jh} function $C_0$,
defined by
\begin{eqnarray}  
& & \hspace*{-.4cm} C_0(s; m) \equiv 
       C_0(p_1, p_2; m, m, m) 
                    \nonumber \\                    
&&  \hspace*{-.4cm} = \!\!\int\! \frac{d^4\! q}{i\pi^2}
                      \frac{1}{(q^2\! - m^2)\left[(q+p_1)^2\! - m^2\right]\left[(q+p_1 +p_2)^2\!-m^2\right]} 
                    \nonumber \\
\end{eqnarray}
where $p_1^2 = p_2^2 = 0$ and $(p_1 + p_2)^2 = s$.
The well-known top loop 
is~\cite{HUTP-77/A084,toploop}
\begin{equation}
\tau_t A_t(\tau_t) = 
 - 4 M_t^2 (1-\tau_t) C_0(4 M_t^2 \tau_t; M_t) - 2 \; .
\end{equation}
Using these expressions, it is straightforward to calculate the
effects of one or more scalars on the gluon fusion rate.

In the limit where the Higgs mass is small, $M_h \ll M_t,M_i$, the 
amplitudes are real, and asymptote to mass-independent values:  
$A_t(0) = -4/3$ and $A_i(0) = -1/3$.  
This yields the following  change in the  $h \ra gg$ width: 
\begin{equation}
\hspace*{-0.2cm}
\left( \frac{\Gamma(h \ra gg) }{\Gamma(h \ra gg)_{\mathrm{SM}} } \right)_{\!\! M_h \ll M_t, M_i}\!\!\! 
\approx  \, \left|1 + \sum_i c_i \kappa_{ii}  
      \frac{v^2}{8 M_i^2}\right|^2 \, ,
\label{eq:widthapprox}
\end{equation}
where $\Gamma(h \ra \!gg)_{\mathrm{SM}}$ is the standard model width.  
This shows that \emph{suppression} of Higgs
production occurs for $\kappa_{ii} < 0$.\footnote{This was noted in 
Ref.~\cite{Boughezal:2010ry} in the context of a real scalar color octet, 
but not explored further.}
For a single colored scalar, a substantial 
cancellation between the top and  scalar loops
is possible when its mass is related to its Higgs portal
coupling by $M_i  \approx v \sqrt{c_{ii} |\kappa_i |/8}$. 

In the particular case $M_h = M_i = M_t$, the amplitudes are $A_t(1/4) = -8(1 - \pi^2/12)$ and $A_i(1/4) = -4(\pi^2/9 - 1)$, 
so that
\begin{equation}
\hspace*{-0.1cm} \left(\frac{\Gamma(h \ra gg) }{\Gamma(h \ra \!gg)_{\mathrm{SM}} }\right)_{\!\! M_h = M_i = M_t} \!\!\! \approx 
\left|1 +  \frac{2}{3} \left(\!\frac{\pi^2 - 9 }{12 - \pi^2} \!\right) \sum_i c_i \kappa_{ii}  \right|^2 
\label{eq:widthapprox}
\end{equation}
where we used $v\approx \sqrt{2} M_t$.
Substantial cancellation in this case requires $\sum_i c_i \kappa_{ii} \approx - 3.7$.

Color-octet real (complex) scalars have  $c_i = 3$ (6),  so that the above particular cases show that 
gluon fusion may be strongly suppressed with order-one couplings. We will analyze the color-octets 
in Section III. 
In the case of color-triplet complex scalars, the Casimir is smaller, $c_i = 1$, so that several triplets are necessary to obtain 
substantial suppression with order-one $\kappa_{ii}$ couplings. 

Supersymmetric models automatically have color-triplet scalars with 
substantial couplings to the Higgs sector.  The possibility of
destructive interference between the top loop and loops of stops
has been explored previously \cite{susysuppression}.  
In the supersymmetric case,
the coupling to the Higgs is determined by supersymmetric as well as
supersymmetry-breaking interactions, and so the size and sign of
the contribution is model-dependent.  
In the limit of no supersymmetry breaking with
$\tan\beta = 1$ and $\mu = 0$, there are two mass eigenstates,
a pure $\tilde{t}_L$ and $\tilde{t}_R$ with masses equal to the
top mass and Higgs portal coupling given by $\kappa =  y_t^2$.  In the limit $M_h \ll M_t = M_{\tilde{t}}$, 
the addition of the stops results in an increase in the amplitude 
by a factor of 3/2, and thus an increase in the Higgs production rate 
by a factor of 9/4.  This is indicative of the size of the
correction that colored scalars can provide, but this particular
limit does not yield a realistic model of low-energy supersymmetry 
due to the lack of both tree-level \emph{and} one-loop corrections 
to the Higgs mass itself.

If the Higgs mass is large enough for an on-shell decay to proceed,
the $h \to gg$ amplitude 
develops an imaginary part.  
Two on-shell decays
could occur:  $h \ra t\bar{t}$ and/or $h \ra \phi_i^{(\dagger)}\phi_i$.  
This leaves four distinct possibilities: \\
$i)$  $M_h < 2 M_i, 2 M_t$:  No imaginary part is generated 
for the amplitudes; suppression of gluon fusion arises entirely
through cancellation of the real parts of the diagrams. \\
$ii)$ $2 M_i < M_h < 2 M_t$:  An imaginary part is generated 
for the amplitude involving colored scalars.  It increases rapidly 
(in magnitude), such that 
$\mathrm{Im}[A_i(\tau_i)] = \mathrm{Re}[A_i(\tau_i)]$ 
is achieved already once $M_h \simeq 2.15 M_i$. 
This results in a significant non-cancelable contribution to the 
amplitude for Higgs production through gluon fusion.  \\
$iii)$  $2 M_t < M_h < 2 M_i$:  An imaginary part is generated 
for the amplitude involving top quarks.  It increases more slowly, 
we find 
$\mathrm{Im}[A_t(\tau_t)] = (1/4, 1/2, 1) \times \mathrm{Re}[A_t(\tau_t)]$ 
occurs when $M_h \simeq (2.3, 2.5, 3.1) M_t$.  
Hence, there is a region of parameter space when $2 M_t \lsim M_h$
for which sizeable cancellation in the real parts remains
sufficient to suppress Higgs production through gluon fusion. \\
$iv)$  $2 M_i, 2 M_t < M_h$: Imaginary parts are generated
for both amplitudes involving colored scalars as well as 
top quarks.  Interestingly, for $2 M_i \simeq M_h$ and with
$M_h \gsim 2 M_t$, both the real and imaginary contributions 
to $A_t$ and $A_i$ are negative.  This suggests there is an 
interesting regime where both the real and imaginary parts
of the contributions from top loops and colored scalar loops
can \emph{simultaneously} destructively interfere. \\
We will see all four of these cases arise in the specific model
involving a color-octet scalar considered in the next section.

It is also interesting to estimate how small the $gg \to h$ rate could
be made in principle.  Note that so far we
have neglected other quark contributions to the amplitude.
While it is possible for scalar loop contributions to cancel
the sum of the real parts of the top loop and the much smaller
light quark contributions, without extraordinary tuning 
it is not possible to also cancel the 
small \emph{imaginary} part that accompanies $h \ra b\bar{b}$.  
Even for the smallest Higgs mass allowed by LEPII,  
we find the absolute value of the imaginary part of the $b$-quark loop 
contribution is smaller than 10\% of the absolute value 
(real part) of the top quark contribution to the amplitude.
Hence, the gluon fusion Higgs production rate could be as small as 1\% of the standard model rate 
while not running afoul of this lower bound.

\section{Color-octet real scalar}
\label{sec:octet}
\setcounter{equation}{0}

Let us now consider the standard model plus an electroweak-singlet, color-octet real 
scalar field $\octet^a$. The most general renormalizable Lagrangian involving $\octet^a$ is
\begin{eqnarray}
\mathcal L_\octet &=&   
    \frac{1}{2}(D_{\mu}\octet^a)^2 
    - \frac{1}{2} \left( M^2_0 + \kappa \, H^{\dag}H \right) \octet^a \octet^a 
    \nonumber \\ [2mm]
& &{} - \mu_\octet \, d_{abc} \octet^a \octet^b \octet^c  
      - \frac{\lambda_{\octet}}{8}(\octet^a\octet^a)^2 \nonumber \\
& &{} - \lambda'_{\octet} d_{abe} d_{cde} 
                          \octet^a\octet^b\octet^c\octet^d \; .
\label{eq:octetlagrangian}
\end{eqnarray}
Here $\kappa$,\,$\lambda_\octet$ and $\lambda'_\octet$ are dimensionless real parameters, 
$M_0$ and $\mu_\octet$ are real parameters of mass dimension +1,
and $d_{abc}$ is the totally symmetric $SU(3)$ tensor.
After electroweak symmetry breaking, the octet obtains the physical mass
\begin{equation}
M_\octet^2 = M_0^2 + \frac{\kappa}{2} v^2 \; ,
\end{equation}
which we require to be positive definite.  This implies a constraint
on the bare (mass)$^2$, $M_0^2 > - \kappa v^2/2$.
As we saw in Sec.~\ref{sec:noggh}, a negative Higgs portal 
interaction, $\kappa < 0$, is interesting because it leads to 
destructive interference between the top-quark and scalar loops.
To ensure that $\octet$ does not acquire a VEV, one needs to impose 
$\lambda_{\octet}> 0$ and $ |\mu_\octet| \lsim M_\octet$ 
(the precise upper limit depends on $M_h$, $\lambda_{\octet}$ and $\lambda'_{\octet}$, 
as well as on the sign of  $\mu_\octet$).

\begin{figure}[t!]
\includegraphics[width=0.45\textwidth]{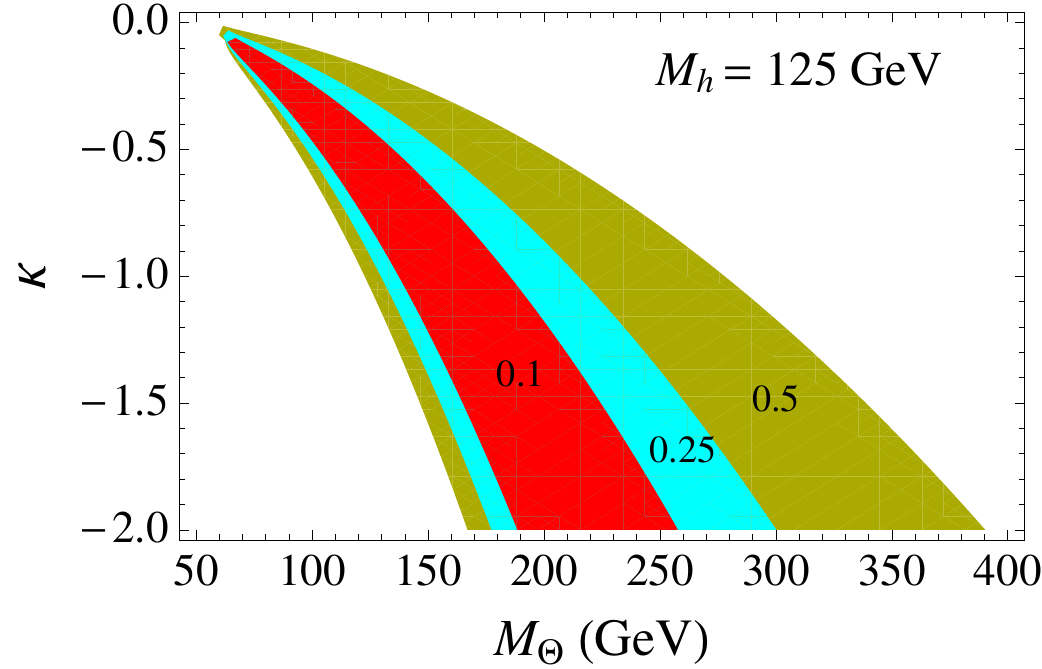} \\ [2mm]
\includegraphics[width=0.45\textwidth]{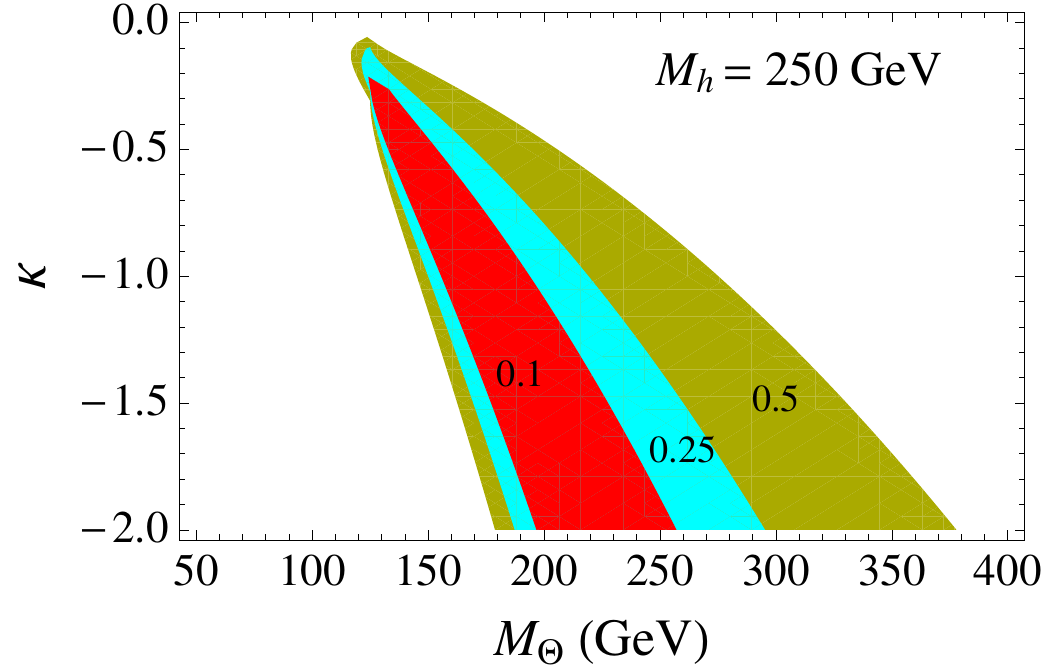} \\ [2mm]
\includegraphics[width=0.45\textwidth]{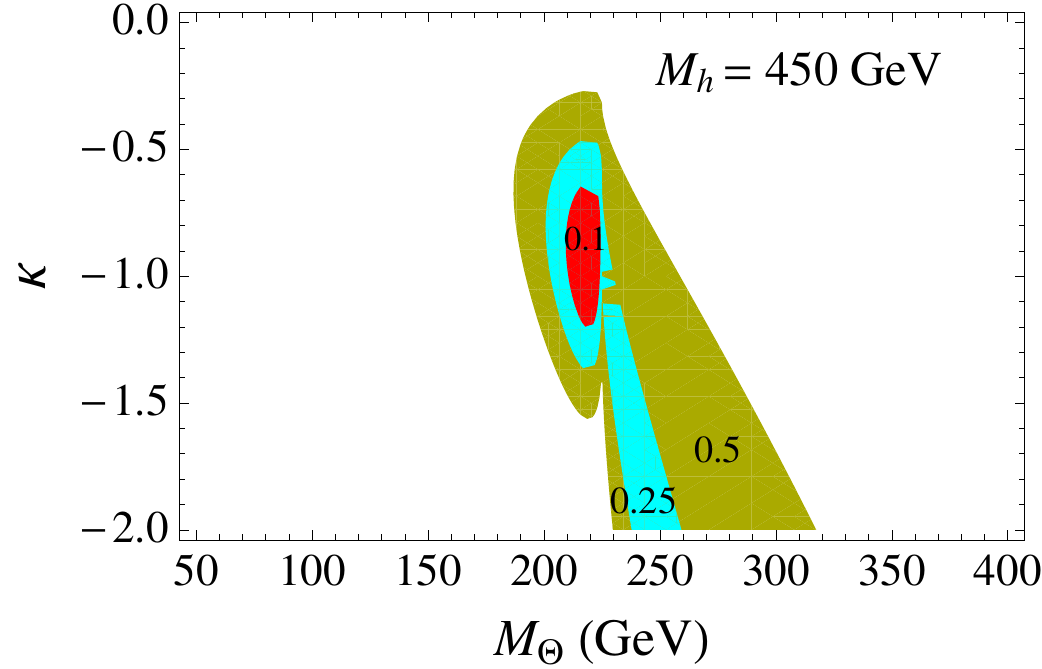}
\caption{Contours of the Higgs production cross section 
through gluon fusion at leading order, including the effects of a color-octet
real scalar having Higgs portal coupling $\kappa$ and mass $M_\octet$, normalized to the standard model value.  
The inner (red), middle (blue), outer (green) regions
correspond to 
$\sigma(pp\rightarrow h)/\sigma(pp\rightarrow h)_{SM} < 0.1, 0.25, 0.5$ 
respectively.  The top, middle, and bottom panels show increasing
Higgs mass.  As thresholds for $h \ra 2$-body decays are 
crossed, qualitative changes in the suppression of the 
Higgs production through gluon fusion are evident.} 
\label{fig:matching}
\end{figure}

The effects of a color-octet scalar on the suppression of 
Higgs production through gluon fusion can be obtained directly 
from the results of Sec.~\ref{sec:noggh}, substituting $c = C_A = 3$.  
We evaluate the Passarino-Veltman function using the LoopTools package \cite{Hahn:1998yk}.
The parameter space is controlled by the Higgs mass and 
two parameters in the octet model, $(M_\octet, \kappa)$. In Fig.~\ref{fig:matching} we show contours of 
$\sigma(gg\rightarrow h)$ in the $M_\octet$ versus $\kappa$ plane 
for three choices of Higgs mass, $M_h = 125,\, 250,\, 450$~GeV\@.  
In this contour plot, we have normalized the cross sections 
to the standard model value at leading order.  Working within the narrow width approximation, 
all parton distribution effects factorize and the ratio of cross sections is simply 
the ratio of widths, $\Gamma(h \rightarrow gg )/\Gamma(h\rightarrow gg)_{\rm SM}$.

The striking result is that the Higgs production is substantially
reduced in a large region of the $(M_\octet, \kappa)$ parameter space.
In the $M_h = 125, 250$~GeV panels, the contours cut off fairly 
rapidly near $M_\octet = M_h/2$, corresponding to when the scalar 
contribution to the amplitude develops an imaginary part 
resulting from $h \ra \octet\octet$ going on-shell.  

In the $M_h = 450$~GeV panel, since the decay $h \ra t\bar{t}$ 
goes on-shell, the amplitude again develops an imaginary part 
from the top loop.
Here we see two regions where suppression to Higgs production
is possible.  The first region, when $M_\octet > M_h/2$, 
is analogous to similar regions for lower Higgs masses.
However, since there is a non-cancelable imaginary part,
the size of the cross section suppression is more limited
within the range of parameters shown.  The second region, when 
$M_\octet \lsim M_h/2$, both the top loop and scalar loops
have both real and imaginary parts that partially 
destructively interfere.  Surprisingly, the interference
can be just as effective in this region of parameter space
as we found when the amplitudes were purely real, 
$M_h < 2 M_t, 2 M_\octet$.

\begin{figure}[t!]
\includegraphics[width=0.45\textwidth]{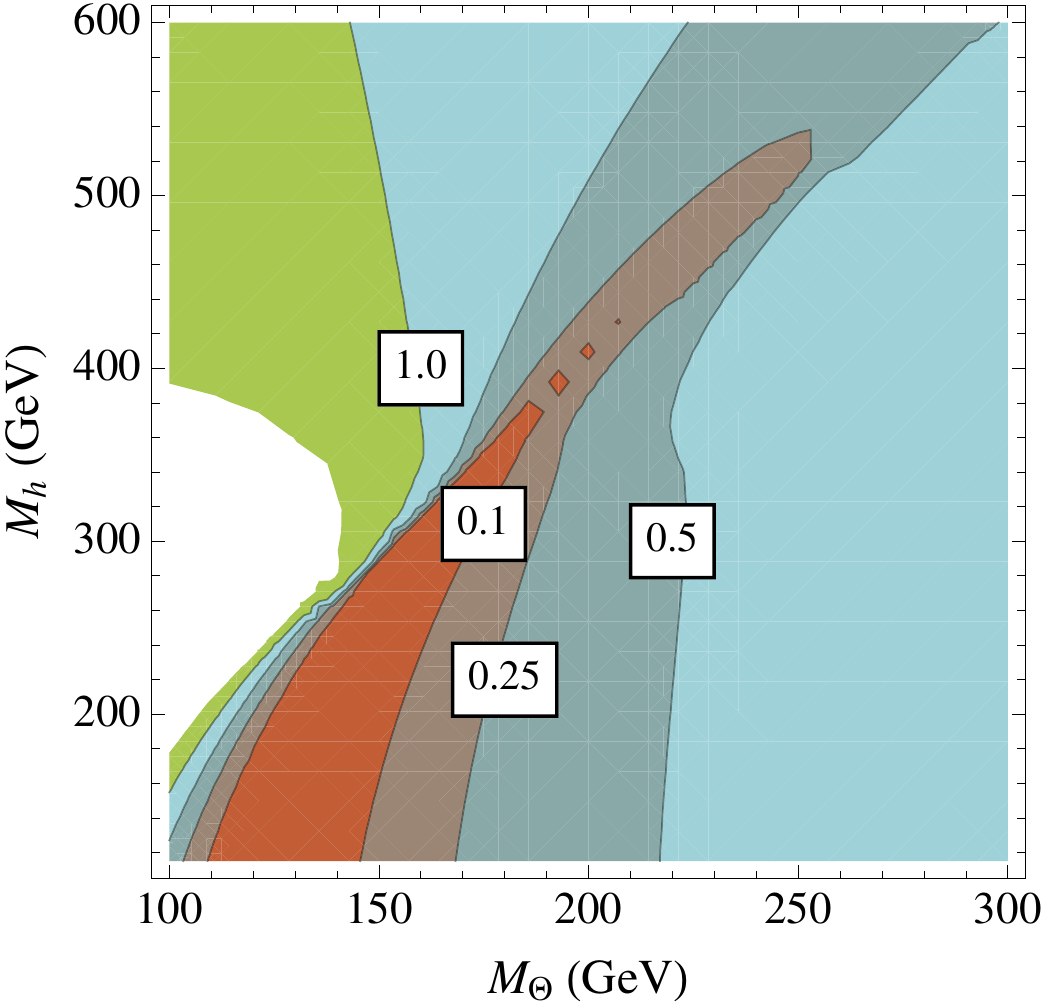} \\ [2mm]
\includegraphics[width=0.45\textwidth]{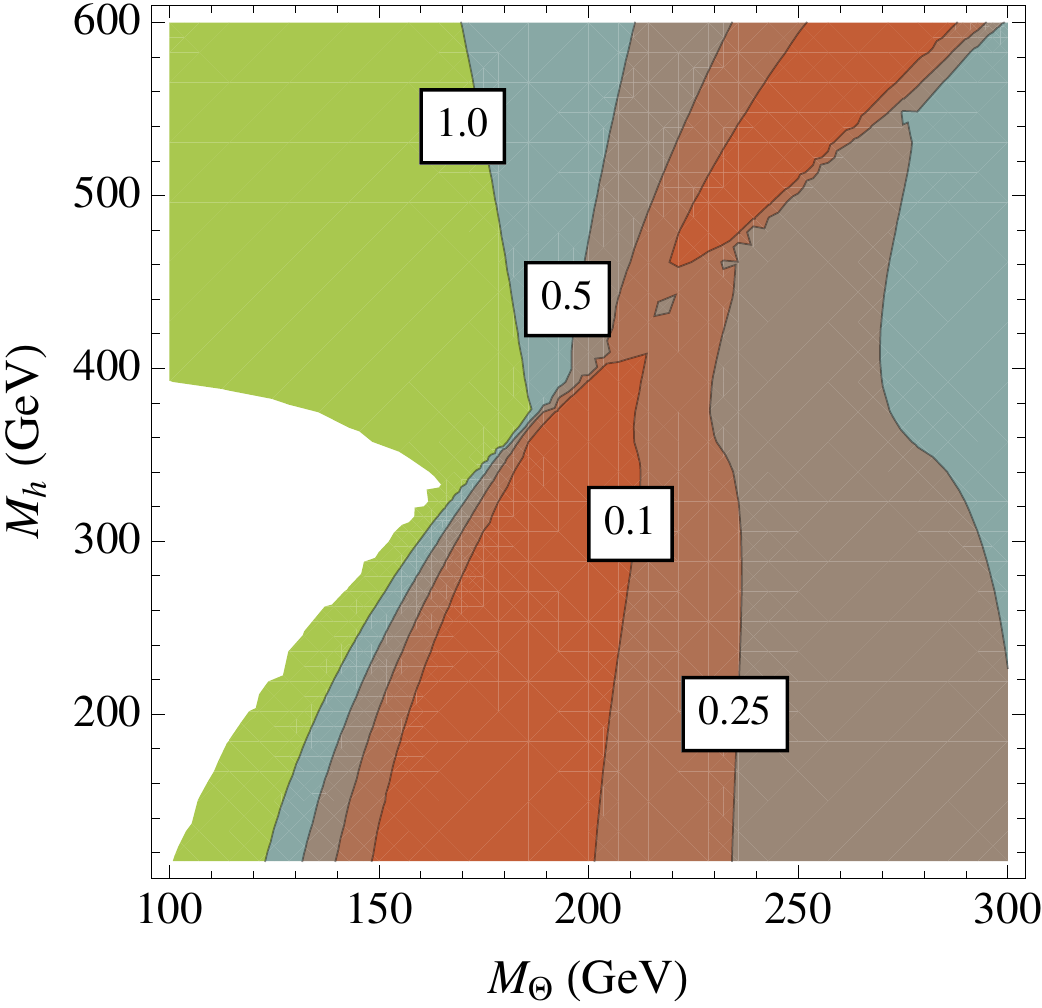}
\caption{Contours of the Higgs production cross section 
through gluon fusion at leading order, including the effects of a color-octet real
scalar, normalized to the standard model value.  
Unlike Fig.~\ref{fig:matching}, we have fixed the Higgs 
portal coupling to $\kappa = -0.6, -1.2$ in the upper and lower plots,
respectively, while allowing $M_h$ and $M_\octet$ to vary.}
\label{fig:kappaplots}
\end{figure}

In Fig.~\ref{fig:kappaplots} we again show contours of 
$\sigma(gg\rightarrow h)$, normalized to the SM value, 
but now in the $M_\octet$ versus $M_h$ plane while holding 
$\kappa = -0.6, -1.2$ fixed at two values.
Much of the structure of the contours is determined by
the threshold for $h \ra \octet \octet$ to go on-shell,
which is the clear diagonal line in the plots satisfying
$M_h = 2 M_\octet$.   
There are two distinct regions of gluon fusion suppression.
The first is when $M_h < 2 M_\octet$ and $M_h \lsim 2 M_t$
in the lower center of both plots.  In this case, the real
parts between the two diagrams are destructively interfering,
even when $h \ra t\bar{t}$ goes (slightly) on-shell, due to 
the slow rise of the top amplitude's imaginary part.  
In the second region, $M_h > 2 M_\octet, 2 M_t$,
more clearly seen in the lower plot of Fig.~\ref{fig:kappaplots} ($\kappa = -1.2$),  
both real and imaginary parts for the top and scalar amplitudes 
are present and destructively interfere.
It is remarkable that such sizable suppression, between a 
factor of 2 to 10 in the rate for gluon fusion, persists 
throughout much of the parameter space of both plots.

\begin{figure}[t!]
\includegraphics[width=0.45\textwidth]{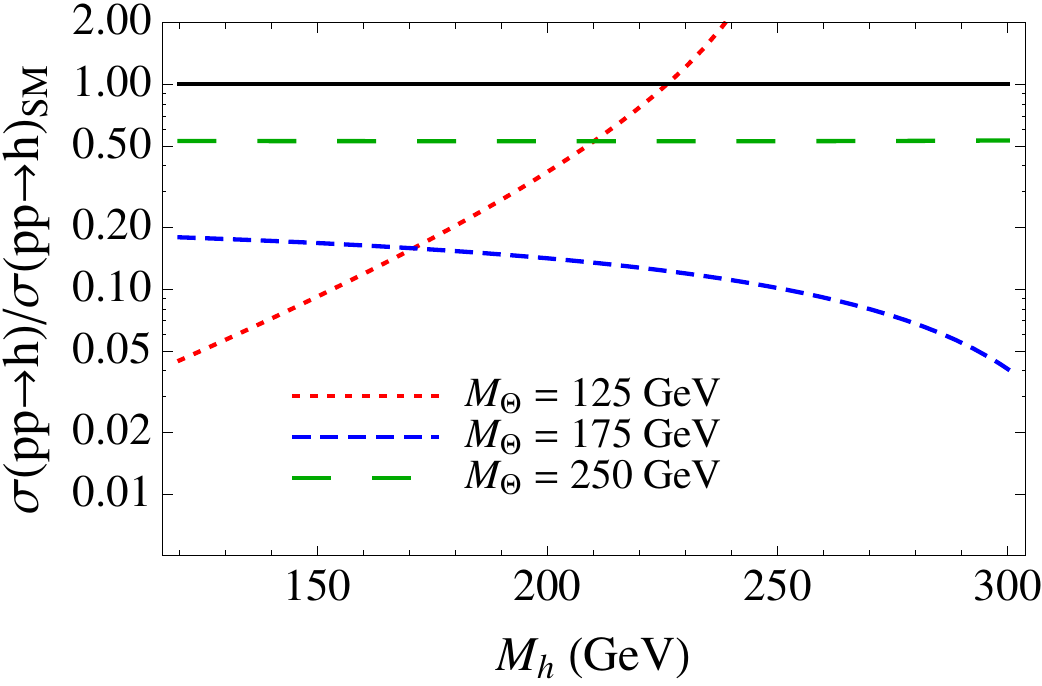}
\caption{Cross section $\sigma(pp \rightarrow h)$ relative to the standard model value for octets of mass $125\,\gev$ (dotted line), $175\,\gev$ (small dashes) and $250\,\gev$ (large dashes) and Higgs portal coupling $\kappa = -0.75$.}
\label{fig:varymh}
\end{figure}

For another perspective, we can fix both the coupling $\kappa$ and the octet mass, then plot the Higgs production cross section as a function of Higgs mass alone. We do this in Fig.~\ref{fig:varymh} for $\kappa = -0.75$ and three different octet masses, $125\,\gev, 175\,\gev$ and $250\,\gev$.

What we see is that Higgs production through gluon fusion
can be suppressed throughout the Higgs mass range 
from the LEP II bound up to largest Higgs masses
that the LHC is currently sensitive to.  We should note 
that our calculations of the cross sections have been performed 
in the narrow Higgs width approximation, and for the largest 
Higgs masses, the finite width effects become increasingly important.

Higgs production through gluon fusion is well-known to have 
large higher-order corrections \cite{higherorder1,higherorder2,arXiv:0709.4227}. 
Extensive higher-order calculations of the effects of a real
scalar color octet on Higgs production were also carried out in 
Ref.~\cite{Boughezal:2010ry}.  These calculations were
applied exclusively to consider \emph{enhancements} in the 
Higgs production rate, and the extent to which they can be 
bounded from data.  We did, however, apply their results to
the negative $\kappa$ region, to estimate the higher order
corrections to the parameter space shown in Fig.~\ref{fig:matching}.
We found that the higher order corrections enhance the
scalar contribution relative to the top loop, and thereby allow for smaller $\kappa$,
by as much as 25\%, holding $M_\octet$ and the Higgs 
cross section fixed.
 
The requirement of a relatively light colored scalar octet
with mass less than a few hundred GeV is obviously of some
concern since it can be copiously produced at the LHC\@.
The signature of the color octet critically depends on its decay.
Given our Lagrangian, Eq.~(\ref{eq:octetlagrangian}), 
the dominant decay is $\octet \to g g$, which proceeds 
at 1-loop through diagrams involving a $\mu_\octet$ vertex and 
$\octet$ running in the loop.  The width for this process is 
very small \cite{Bai:2010dj},
\begin{equation}
\Gamma (\octet \to gg) \approx 
     5 \times 10^{-7} \, \frac{\mu_\octet^2}{M_\octet} \; ,
\end{equation}
but nevertheless leads to prompt decays 
for $\mu_\octet^2/M_\octet > O(10)$ eV\@.

\begin{figure}[t]
\centering
\includegraphics[width=3.2in]{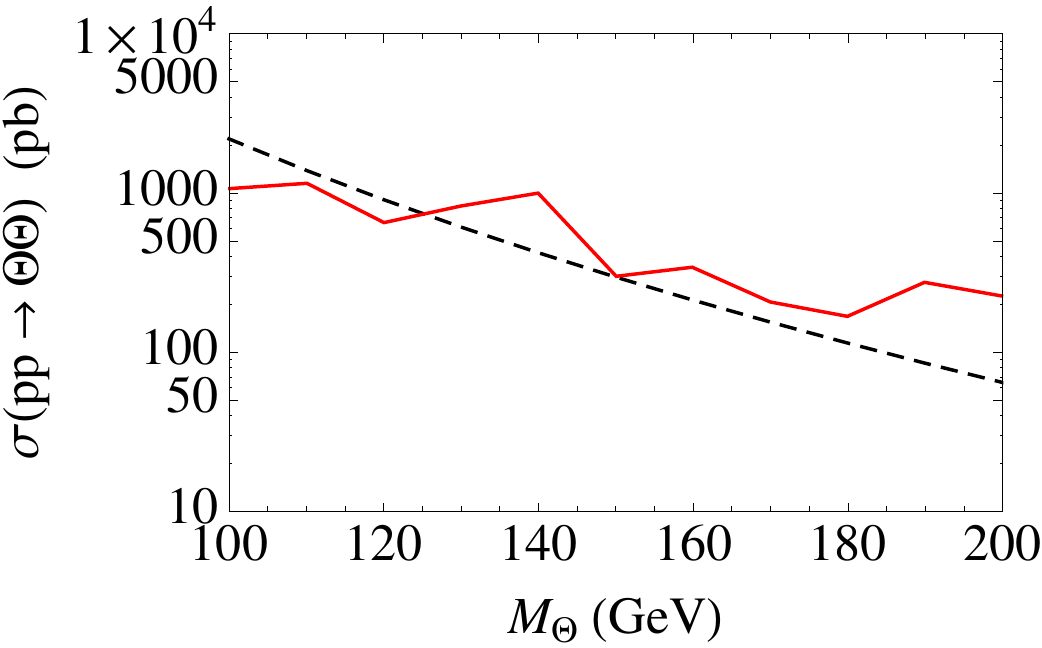}
\caption{Limit on the production cross section for a pair of dijet resonances
from ATLAS~\cite{Aad:2011yh} (solid line), and the leading-order theoretical cross section 
(dashed line)
for pair production of a color-octet real scalar at the 7 TeV LHC\@.}
\label{fig:ssprod}
\end{figure}

The QCD production of color-octet scalars at hadron colliders 
has been studied in various models \cite{HUTP-91-A009,Dobrescu:2007yp,Bai:2010dj,Gerbush:2007fe,Dobrescu:2011px,arXiv:0810.3919}.
Here, $\octet$ production occurs in pairs, 
so that the signature is a pair of dijet resonances 
\cite{HUTP-91-A009, Dobrescu:2007yp,arXiv:0810.3919}.
The cross section at the LHC is large and depends only on 
$M_\octet$ and $\sqrt{s}$ \cite{Dobrescu:2007yp,Bai:2010dj}. 
The ATLAS Collaboration has searched for this signature using the 2010 data \cite{Aad:2011yh}, 
and has set a 95\% CL limit on the cross section shown by the dashed line 
in Fig.~\ref{fig:ssprod}. We also show the leading-order 
theoretical prediction for 
$\octet$ pair production in Fig.~\ref{fig:ssprod}. Comparing these 
two lines we find that the octet real scalar is ruled out for 
$M_\octet$ in the $100$--$125$~GeV range at 95\% CL\@.
The inclusion of next-to-leading order effects would likely increase 
the theoretical cross section, such that a small mass region around 
$150$~GeV is also ruled out.  Note that the production cross section 
for a real scalar is half of that for a 
complex scalar \cite{arXiv:0810.3919}.

Single production of $\octet$ is possible at 1-loop, through gluon fusion, 
and is typically too small to be interesting (the cross section can 
be found in \cite{Gresham:2007ri} for the case of a weak-doublet 
color octet). 

The cancellation we have demonstrated requires the Higgs portal coupling 
to be negative.  The existence of a negative quartic couplings 
suggests we consider the vacuum stability of the full scalar potential. 
At small field values, the requirement of a positive mass squared for 
$\octet$ ensures small fluctuations are stabilized.  
At large field values, we need to consider the other terms in the octet Lagrangian 
(\ref{eq:octetlagrangian}) as well as the Higgs quartic coupling $\lambda_h$.
For simplicity, let us assume that $\lambda_\octet^\prime$ and $\mu_\octet$
are too small to affect the minimization of the potential (this is easily consistent 
with the $\mu_\octet \gtrsim 1$ MeV limit required by prompt $\Theta$ decays). 
The same-field quartics $\lambda_h$, $\lambda_{\octet}$ are positive, 
and so stabilize the large $H$ and large $\octet$ 
directions of field space.  However, negative $\kappa$ could provide 
a direction with a minima lower than the electroweak symmetry 
breaking minimum.  Positive definiteness of the potential 
at large field values is automatic if the potential can be
written in the form $(\sqrt{\lambda_h}\,H^\dagger H - \sqrt{\lambda_\octet}\, \octet^a \octet^a)^2$ plus terms
that are positive definite.  This yields the tree-level 
constraint \cite{Boughezal:2010ry} 
\begin{equation}
|\kappa| < 2 \sqrt{\lambda_{\octet}\,\lambda_h},
\label{eq:franklimit}
\end{equation}
which would appear to somewhat constrict the parameter space of 
our color-octet model.  However, to properly bound $\kappa$, we 
must consider the effects of renormalization group (RG) running on
the couplings in the potential. Evolving to higher energies, 
the quartic coupling $\lambda_{\octet}$ increases.  
This increase happens fairly quickly, driven primarily by the 
$\lambda^2_{\octet}$ term in the beta function, and is enhanced 
by color combinatorial factors.  Equally important, the Higgs 
portal coupling {\em decreases} in magnitude as we go to 
higher energy; $\beta(\kappa) \propto \kappa^2$, so an initially 
large negative $\kappa$ rapidly evolves to a small negative $\kappa$. 
Hence, there is a considerably larger range of $\kappa$ and $\lambda_{\octet}$ 
satisfying the constraint of no deeper minimum in the RG-improved 
effective potential.  We leave a detailed study to future work.

\section{Discussion}
\label{sec:discussion}

The gluon fusion-induced single Higgs production rate at the LHC 
could be substantially suppressed when the standard model 
is extended to include a colored scalar sector that interferes 
destructively with the top-quark loop.  The general class of models
consist of one or more colored scalars with mass less than a 
few hundred GeV\@.  Large suppression of the gluon fusion rate 
is possible throughout the Higgs mass range while having negligible 
effect on the Higgs branching ratios, effectively allowing the 
Higgs boson to exist at any mass given the current LHC limits.

In this paper we have concentrated on a specific model consisting of 
a color-octet real scalar with a negative Higgs portal coupling.
Based on Fig.~\ref{fig:matching}, we find that the interesting
range of color octet masses giving substantial gluon fusion
suppression is roughly $60 \lsim M_\octet \lsim 300$~GeV\@.
In the presence of the cubic coupling given in Eq.~(\ref{eq:octetlagrangian})
the color octets decay to a pair of gluons. Only ATLAS has provided experimental constraints
that impact the model, ruling out the region $100$--$125$~GeV 
to 95\% CL \cite{Aad:2011yh}.  Masses above $125$~GeV are 
allowed by current bounds.  We are not aware of a robust
constraint that rules out the region $60 \lsim M_\octet < 100$~GeV, 
suggesting a more detailed analysis of the viability (or lack thereof) 
of this region would be interesting for experiments to carry out.

It is interesting to correlate the suppression in single Higgs 
production with changes in di-Higgs production.
The set of diagrams contributing to di-Higgs production
consist of both order $\kappa$ (e.g. triangle diagrams) 
as well as $\kappa^2$ (e.g. box diagrams) contributions to the 
amplitude.  When single Higgs production is suppressed, 
the order $\kappa$ diagrams are suppressed.  However, 
larger $|\kappa|$ implies the second class of diagrams
proportional to $\kappa^2$ remain, and are dramatically 
enhanced.  For the color-octet scalar model, we find 
the increased di-Higgs production rate between a factor 
of a few to over $100$ times the SM rate for the same 
Higgs mass \cite{workinprogress}.
An increase in di-Higgs production can also be found 
in the presence of cutoff scale operators \cite{gghhdim6}.

We must emphasize that our analysis of Higgs suppression from 
a single color-octet scalar is merely one model of a large
class of colored scalar models.  The signals of
any given model can be completely different.  For example, 
supersymmetric models with light top squarks can easily have
an order one negative $\kappa$, and yet the canonical search
strategy for stops involves missing energy (when $R$-parity
is conserved) with detailed considerations of stop decay.
A model in which the colored scalars are ``quirks''
bound by a new strongly-coupled sector would yield 
completely different signals (e.g. ~\cite{arXiv:1106.2569}).
Thus, while the searches for specific colored scalars are 
very important, what is more important for Higgs physics
is to study the extent to which the Higgs boson can be observed 
in other channels.

Other Higgs production sources continue to provide
a smaller but non-negligible source for single Higgs signals.
Specifically, associated production ($W h$ and $Z h$) and 
vector-boson-fusion (VBF) production sources remain unchanged.  
The VBF process provides a non-negligible single Higgs production rate
throughout the Higgs mass range, though the rate is roughly a factor 
of 10 smaller than gluon fusion rate for $m_h < 2 m_t$.  
However, existing LHC search strategies 
have been optimized for a gluon-fusion source, and so as far as
we understand, the present Higgs production rate bounds cannot be 
trivially rescaled.  
For example, current search strategies involving the 
$h \ra WW \ra \ell^+\ell^- + \slashchar{E}_T$ final state allow 
$0, 1$ additional jets in the signal \cite{atlas-conf-2011-134}.
The VBF process generically produces two (forward) jets.
Hence, we suspect 
that the current Higgs searches are sensitive to only
a small fraction of the VBF rate.  A more complete study
of how effective the LHC experiments are sensitive to VBF 
production would be really useful. 

In addition, some strategies to constrain the light Higgs mass region
also depend on a convolution of the gluon fusion rate with 
other Higgs production sources. 
For example, the inclusive selection at CMS \cite{cms-pas-hig-11-020}
for the $h \ra \tau\tau$ mode receives a substantial 
contribution from gluon fusion as well as VBF\@. 
Obtaining bounds on the Higgs production cross section in the
presence of light colored scalars therefore requires 
separating out the various sources of Higgs production.

Finally, one new single Higgs production channel is possible:
associated production with a pair of scalars $\phi\phi h$.  
This has been considered before in supersymmetric models 
\cite{susystopstoph}.
For larger $|\kappa|$ and smaller $M_\octet$, this process
can be considerably larger than the similar standard model
process, $t \bar{t} h$~\cite{workinprogress}.  
It would provide the direct 
confirmation that colored scalars are indeed interacting
with the Higgs through the Higgs portal couplings, and thus
responsible for modifying the Higgs production rate. \\

\emph{Note added:}~  Ref.~\cite{Bai:2011aa} also considers
the suppression of Higgs production through colored scalars.

\smallskip

\section*{Acknowledgments}

We thank Spencer Chang, Patrick Fox, Howard Georgi, Arjun Menon, 
Roni Harnik, Martin Schmaltz, and Gerben Stavenga, 
for useful comments and conversations. 
GDK was supported by a Ben Lee Fellowship from Fermilab and 
in part by the US Department of Energy under contract number 
DE-FG02-96ER40969.
Fermilab is operated by Fermi Research Alliance, 
LLC under contract number DE-AC02-07CH11359 with the 
US Department of Energy.



\end{document}